\documentclass[pra,twocolumn,nofootinbib,floatfix,10pt]{revtex4-2}
\usepackage{textcomp,mathcomp}
\usepackage{amsmath}
\usepackage{amssymb}
\usepackage{wasysym}
\usepackage{graphicx}
\usepackage{color,soul}
\usepackage{physics}
\usepackage{siunitx}
\usepackage{dsfont}
\usepackage{float}
\usepackage[english]{babel}
\usepackage{blindtext}
\usepackage[english,nomargin,inline,marginclue,draft]{fixme}
\pdfpageheight\paperheight
\pdfpagewidth\paperwidth

\usepackage[colorlinks,linkcolor=blue,anchorcolor=blue,citecolor=blue,urlcolor=blue]{hyperref}

\fxusetheme{colorsig}
\FXRegisterAuthor{cg}{acg}{CG}  
\FXRegisterAuthor{th}{ath}{\color{blue}TH}  
\FXRegisterAuthor{ib}{aib}{\color{red}IB} 
\FXRegisterAuthor{sh}{ash}{\color{cyan}SH} 
\FXRegisterAuthor{db}{adb}{\color{green}DB} 
\FXRegisterAuthor{ps}{aps}{PS}
\makeatletter
\renewcommand*\FXLayoutInline[3]{%
  {\@fxuseface{inline}\ignorespaces{\color{fx#1}[#3: #2]}}}
\makeatother

\long\def\symbolfootnote[#1]#2{\begingroup%
\def\thefootnote{\fnsymbol{footnote}}\footnotetext[#1]{#2}\endgroup}

\def\nobreakbefore{%
  \relax\ifvmode\else
    \ifhmode
      \ifdim\lastskip > 0pt\relax
        \unskip\nobreakspace
      \else 
        \nobreakspace
      \fi
    \fi
  \fi
}
\let\oldcite\cite
\renewcommand\cite{\nobreakbefore\oldcite}

\begin{document}
\title{Observation of $s$-$p$ Coupling in Ultralong-Range Rydberg Molecules }

\author{Qing Li$^{1,2}$}
\thanks{These authors contributed equally to this work.}
\author{Shi-Yao Shao$^{1,2}$}
\thanks{These authors contributed equally to this work.}
\author{Jun Zhang$^{1,2,3}$}
\author{Han-Chao Chen$^{1,2,3}$}
\author{Li-Hua Zhang$^{1,2,4}$}
\author{Bang Liu$^{1,2,4}$}
\author{Guang-Can Guo$^{1,2,3}$}
\author{Dong-Sheng Ding$^{1,2,3,4,\textcolor{blue}{\dagger}}$}
\author{Bao-Sen Shi$^{1,2,3}$}

\affiliation{$^1$Laboratory of Quantum Information, University of Science and Technology of China, Hefei, Anhui 230026, China.}
\affiliation{$^2$Anhui Province Key Laboratory of Quantum Network, University of Science and Technology of China, Hefei 230026, China.}
\affiliation{$^3$Hefei National Laboratory, University of Science and Technology of China, Hefei 230088, China}
\affiliation{$^4$Davinci Quantum Technology Co., Ltd, Hefei, Anhui 230026, China}

\date{\today}
\symbolfootnote[2]{dds@ustc.edu.cn}

\maketitle

\textbf{Ultralong-range Rydberg molecules, formed by the interaction between a highly excited Rydberg atom and a ground-state atom, provide a unique platform for exploring quantum phenomena spanning nanometer- to micrometer distances as well as exotic few-body interactions. The molecular binding arises from electron-atom scattering, which is conventionally described by an independent \(s\)- and \(p\)-wave channels picture. Here we report the experimental observation of \(s\)-\(p\) coupling in \(\text{Rb}(nS)\)-Rb(\(5S\)) ultralong-range Rydberg molecules via high-resolution photoassociation spectroscopy. By varying the principal quantum number \(n\) from 31 to 36, we resolve an avoided crossing between the \(s\)- and \(p\)-wave molecular states: whenever a vibrational level bound by the \(p\)-wave well approaches one bound by the \(s\)-wave well in energy, the two states are mixed by the off-diagonal coupling \(U_{\mathrm{sp}}\), and the shifted \(s\)-wave level is identified, which is consistent with theoretical predictions. This mixing is most pronounced for low-node \(p\)-wave molecules, whose binding energies lie closer to those of the \(s\)-wave molecules. The resulting level repulsion and redistribution of spectral intensity reveal the mixed \(s\)-\(p\) character of ultralong-range Rydberg molecules and provide a more precise molecular spectroscopy.}

\section*{INTRODUCTION}
In recent years, various kinds of Rydberg molecules have been extensively studied \cite{greene2000creation,bendkowsky2009observation,2002Macrodimers,2009ObservationOverstreet,2010Giant,2021Rydbergatomion,Zuber2021ObservationOA,zou2023observation}. Among these, ultralong-range Rydberg molecules, which consist of a single highly excited (Rydberg) atom and a neutral ground-state atom, have emerged as a versatile testbed for investigating quantum many-body and few-body phenomena over an extraordinary range of length scales---from the sub-nanometer size of the atomic core to the micrometer-scale extent of the Rydberg electron orbit  \cite{greene2000creation,shaffer2018ultracold,bendkowsky2009observation}. Such molecules are bound by the scattering of the slow Rydberg electron off the ground-state perturber, a mechanism that is fundamentally distinct from the covalent or ionic bonds of conventional diatomic species. The large polarizability and permanent electric dipole moment (with thousands of Debye) of ultralong-range Rydberg molecules give rise to unique features that are fundamentally absent in conventional molecules \cite{kamenski2014formal,li2011homonuclear,booth2015production,niederprum2016observation,kleinbach2017photoassociation,althon2023exploring,2020Cesium,bai2023dissociation}. The spatial orientation of such molecules can be manipulated via external fields, enabling precise control of their alignment \cite{krupp2014alignment,hummel2019alignment}. In the simplest homonuclear case, the molecular potential arises from electron-atom scattering described by the Fermi pseudopotential \cite{fermi1934sopra, omont1977theory}, yielding oscillatory potential curves tied to the Rydberg electron's density modulation \cite{greene2000creation}. Early experiments on alkali metal atoms established that the bound-state spectrum varies with the principal quantum number \(n\), a trend attributed to the slowly varying scattering lengths and wavefunction scaling  \cite{bendkowsky2009observation,bendkowsky2010rydberg,tallant2012observation,bellos2013excitation,gaj2014molecular,krupp2014alignment,anderson2014photoassociation,gonzalez2015rotational,desalvo2015ultra,sassmannshausen2015experimental,eiles2015ultracold}. In parallel, the nodal structure and spectroscopic signatures of \(p\)-wave molecules were found to display a rich variety of features distinct from those of \(s\)-wave molecules  \cite{engel2019precision,legrand2025revealing,giannakeas2020generalized,fey2019effective, eiles2017hamiltonian,niederprum2016observation,anderson2014angular,bendkowsky2010rydberg,exner2025observation,2025Nonadiabatic}, raising the question of how the two channels interact when their potentials become energetically close.

\begin{figure*}[t] 
    \centering \includegraphics[width=1\textwidth]{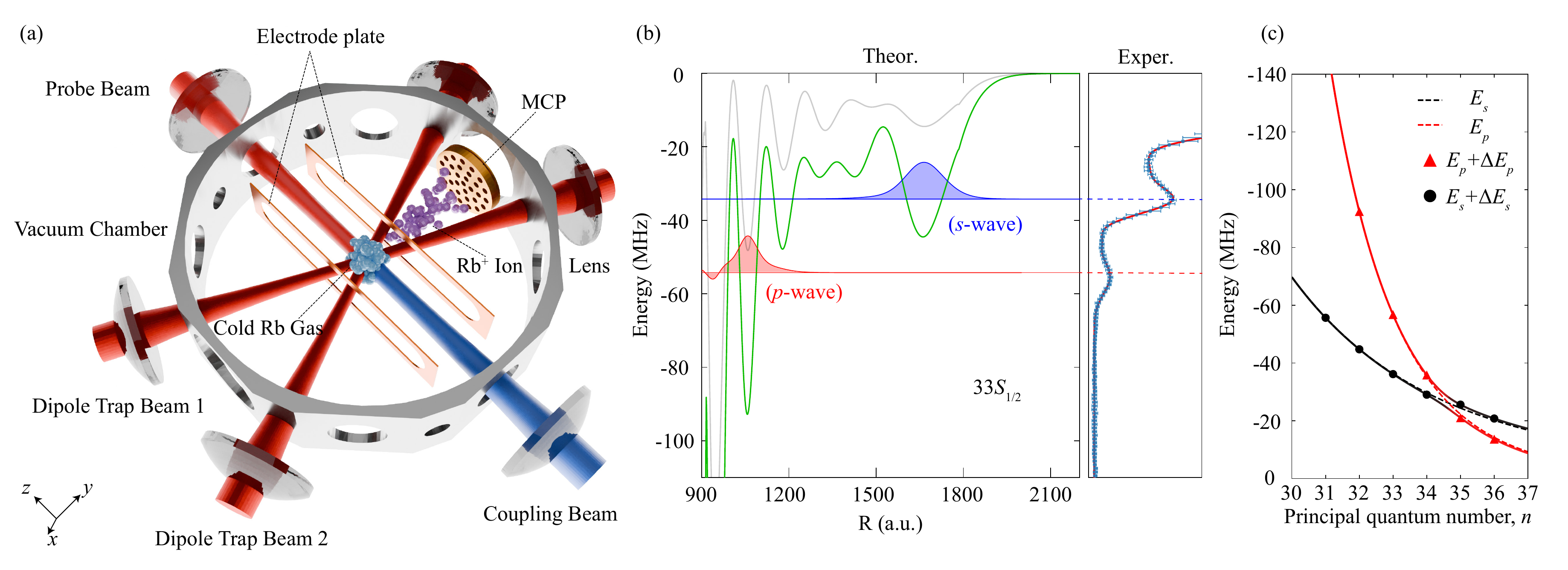} 
    \caption{ \textbf{Excitation scheme and molecular spectra of long-range Rydberg molecules.} (a) Experimental setup. The cold Rb gas is confined in an optical dipole trap formed by two dipole trap beams. Subsequently, the atoms are excited by counter-propagating probe and coupling laser beams through a two-photon excitation process. The electrode plates are used both to apply an external electric field during excitation and to accelerate the $\mathrm{Rb}^+$ ions generated from Rydberg atoms or molecules toward the MCP detector for ion-signal detection. (b) ${}^3\Sigma(5s\text{-}33s)(\nu=0)$ molecular spectra. The left panel shows the theoretically calculated adiabatic potential energy curve for $33S_{1/2}$, where the green curve corresponds to the triplet states and the gray curve corresponds to the mixed singlet-triplet states, while the right panel presents the experimentally measured molecular spectra. The energy reference is set to the corresponding Rydberg atomic line. The outer shallow potential wells of triplet states correspond to molecular vibrational states predominantly formed by $s$-wave scattering, whereas the inner deep potential wells are mainly dominated by $p$-wave scattering. The red curve in right panel represents multi-Gaussian fits to the averaged data. Error bars represent statistical uncertainties derived from 8 independent measurements. (c) Theoretically calculated vibrational energy levels of several s-wave (black circles) and p-wave (red triangles) states for different principal quantum numbers are shown. The red and black solid lines indicate the avoided crossings between them. In the theory, the $s$-$p$ coupling strength is set as $U_{\text{sp}}=2$ MHz.} 
    \label{fig.1} 
\end{figure*}

In the \(s\)-wave limit, the Rydberg molecular potential is proportional to the electron probability density, which scales as \(|\psi|^2 \propto n^{-6}\) for a normalized state whose wavefunction amplitude scales as \(\psi \propto n^{-3}\). However, as the principal quantum number \(n\) varies, the radial nodes of the electron wavefunction shift through space, modulating the local electron density sampled by the ground-state perturber. This nodal modulation critically affects the \(p\)-wave scattering interaction, which likewise derives from the gradient term \cite{gallagher1994rydberg,shaffer2018ultracold,fey2020ultralong}. Moreover, the energy-dependent \(p\)-wave scattering length, which enters the Fermi pseudopotential through a term proportional to the gradient of the electron wavefunction, can exhibit resonances and sign changes as a function of electron collision energy---and thus of \(n\) \cite{hotop2003resonance,sassmannshausen2016long,maclennan2019deeply,exner2025observation}. These effects make the \(s\)- and \(p\)-wave molecular potentials move relative to each other as \(n\) varies. The coupling between the \(s\)- and \(p\)-wave scattering channels becomes significant when the energies of these two kinds of molecular potentials approach each other. In this case, the interplay between wavefunction phase and scattering resonance could, in principle, produce a strong selectivity in deeply bound \(p\)-wave molecular states, and further investigation into this coupling effect is required. 

In this work, we report the experimental observation of \(s\)-\(p\) coupling in \(\mathrm{Rb}(nS)\)-Rb(\(5S\)) ultralong-range Rydberg molecules via high-resolution photoassociation spectroscopy. By varying the principal quantum number \(n\) from 31 to 36, we resolve an avoided crossing between the \(s\)- and \(p\)-wave molecular energies: whenever a \(p\)-wave vibrational level approaches an \(s\)-wave level, the two states are mixed by \(U_{\mathrm{sp}}\), and the \(s\)-wave level is shifted depending on the relative magnitude of $E_s$ and $E_p$. This mixing is most pronounced for low-node \(p\)-wave molecules, whose binding energies lie closer to those of the \(s\)-wave molecules. Incorporating the $s$-$p$ coupling thus provides a more refined and accurate description of the molecular spectrum.

\section*{Results}
\begin{figure*}[t]
    \centering
    \includegraphics[width=1\textwidth]{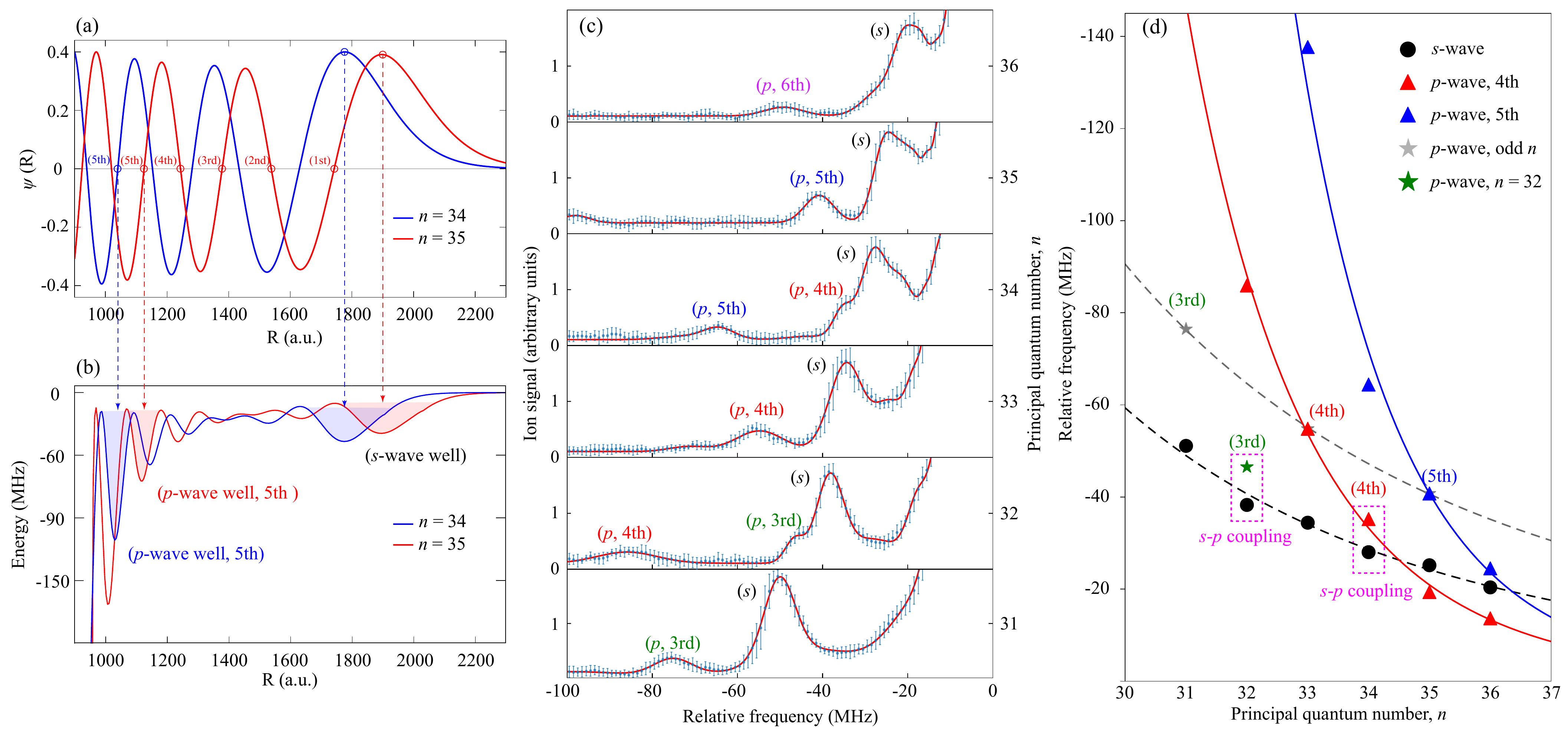}
    \caption{\textbf{Measurement of node-dependent Rydberg molecular states}  (a) Radial wavefunctions of the Rydberg electron for principal quantum numbers $n=34$ (blue) and $n=35$ (red). The first five nodes of the $n=35$ wavefunction and the fifth node of the $n=34$ wavefunction are marked by circles. The different positions of the fifth nodes for different principal quantum numbers lead to distinct depths of the corresponding $p$-wave potential wells. (b) Calculated adiabatic potential energy curves for Rydberg molecules with $n=34$ (blue) and $n=35$ (red). The potential wells dominated by $s$-wave and $p$-wave interactions are distinguished by their internuclear separations $R$. The $p$-wave wells for $n=34$ and $n=35$ are highlighted by blue and red shaded regions, respectively, originating from the fifth wavefunction nodes shown in panel (a). (c) Measured Rydberg molecular spectra for $n=31$ to $36$. The blue dots represent the averaged experimental data, while the red curves are multi-Gaussian fits. The observed resonances are labeled as $(s)$ and ($p$, $i$-$\mathrm{th})$, corresponding to molecular states dominated by $s$-wave and $p$-wave interactions, respectively, where $i=3$--$6$ denotes the nodal order of the corresponding $p$-wave potential well. Error bars represent statistical uncertainties obtained from 8 independent measurements. (d) Resonance positions of the $s$-wave molecular states (black) and the $p$-wave molecular states (red) associated with the fourth wavefunction nodes, plotted as functions of the principal quantum number $n$. The dashed line is fitted curve for the $i$-th node ($i$ = 3, 4, 5) $p$-wave molecule signals. Power-law fits yield scaling exponents of $n^{*-5.25}$ (black), $n^{*-4.69}$ (gray) ,$n^{*-14.54}$ (red), and $n^{*-17.69}$ (blue), respectively. The purple dashed box represents the $s$-$p$ coupling, and the $s$-wave level is obviously shifted from the fitted curve.
 }
    \label{fig.2}
\end{figure*}

\subsection*{Two-channel model and avoided crossing}
The interaction between a Rydberg atom in an \(nS\) state and a ground-state atom of the same species can be described, in the low-energy limit, by the Fermi pseudopotential \cite{fermi1934sopra}. The resulting molecular potential as a function of internuclear distance \(R\) consists of two parts. The \(s\)-wave contribution is
\begin{equation}
V_s(R) = \frac{2\pi\hbar^2}{\mu} a_s(E)\,|\psi_{ns}(R)|^2,
\label{eq:Vs}
\tag{1}
\end{equation}
and the \(p\)-wave contribution is
\begin{equation}
V_p(R) = \frac{6\pi\hbar^2}{\mu} a_p^3(E)\,|\nabla \psi_{ns}(R)|^2,
\label{eq:Vp}
\tag{2}
\end{equation}
where \(\mu\) is the electron mass, \(a_s(E)\) and \(a_p(E)\) are the energy-dependent \(s\)- and \(p\)-wave scattering lengths, respectively, and \(\psi_{ns}(R)\) is the Rydberg electron wavefunction evaluated at the position of the ground-state atom. Solving the vibrational Schrödinger equation with \(V_s(R)+V_p(R)\) yields the \(s\)-wave vibrational energies \(E_s\) and \(p\)-wave vibrational energies \(E_p\).

By employing a two-photon excitation scheme in a cold atomic ensemble, we can investigate the $s$- and $p$-wave ${}^3\Sigma(5s\text{-}ns)$ molecules, the experimental setup is depicted in Figs.~\ref{fig.1}(a), more details about energy levels, experimental setup and measurement procedure are provided in supplementary materials and Methods section. In Fig.~\ref{fig.1}(b), the left panel shows the theoretical calculation by diagonalizing the Hamiltonian, see more details about the theory in Method sections. The right panel in Fig.~\ref{fig.1}(b) displays the measured molecular spectrum in which the $s$- and $p$-wave ${}^3\Sigma(5s\text{-}35s)$ molecules are clearly resolved at energy of $ E_s$ = -34.42 MHz and $ E_p$ = -54.81 MHz. 

The measured molecular spectra reveal two distinct types of resonances: a shallow \(s\)-wave dimer state appearing at small detuning for all measured \(n\), and a deeply bound \(p\)-wave state whose position depends sensitively on \(n\). As \(n\) varies, the \(p\)-wave resonance moves relative to the \(s\)-wave one, and at certain values of \(n\) the two levels approach each other. In a single-channel picture, the two potentials are independent and their vibrational levels would cross rather than repel. However, the off-diagonal interaction between the multiple electron–atom scattering channels results in $s$-$p$ coupling. This coupling is affected by the nodal structure of the Rydberg electron wavefunction: as \(n\) increases, the radial nodes of \(\psi_{ns}(R)\) shift outward, modulating both the local density \(|\psi_{ns}(R)|^2\) and the local gradient \(|\nabla\psi_{ns}(R)|^2\) sampled by the ground-state atom. Since the \(s\)-wave potential is proportional to the density and the \(p\)-wave potential to the gradient, the two potentials move relative to each other and can become energetically proximate for certain \(n\), which is the prerequisite for strong coupling.

\begin{figure*}[t]
    \centering 
    \includegraphics[width=1\textwidth]{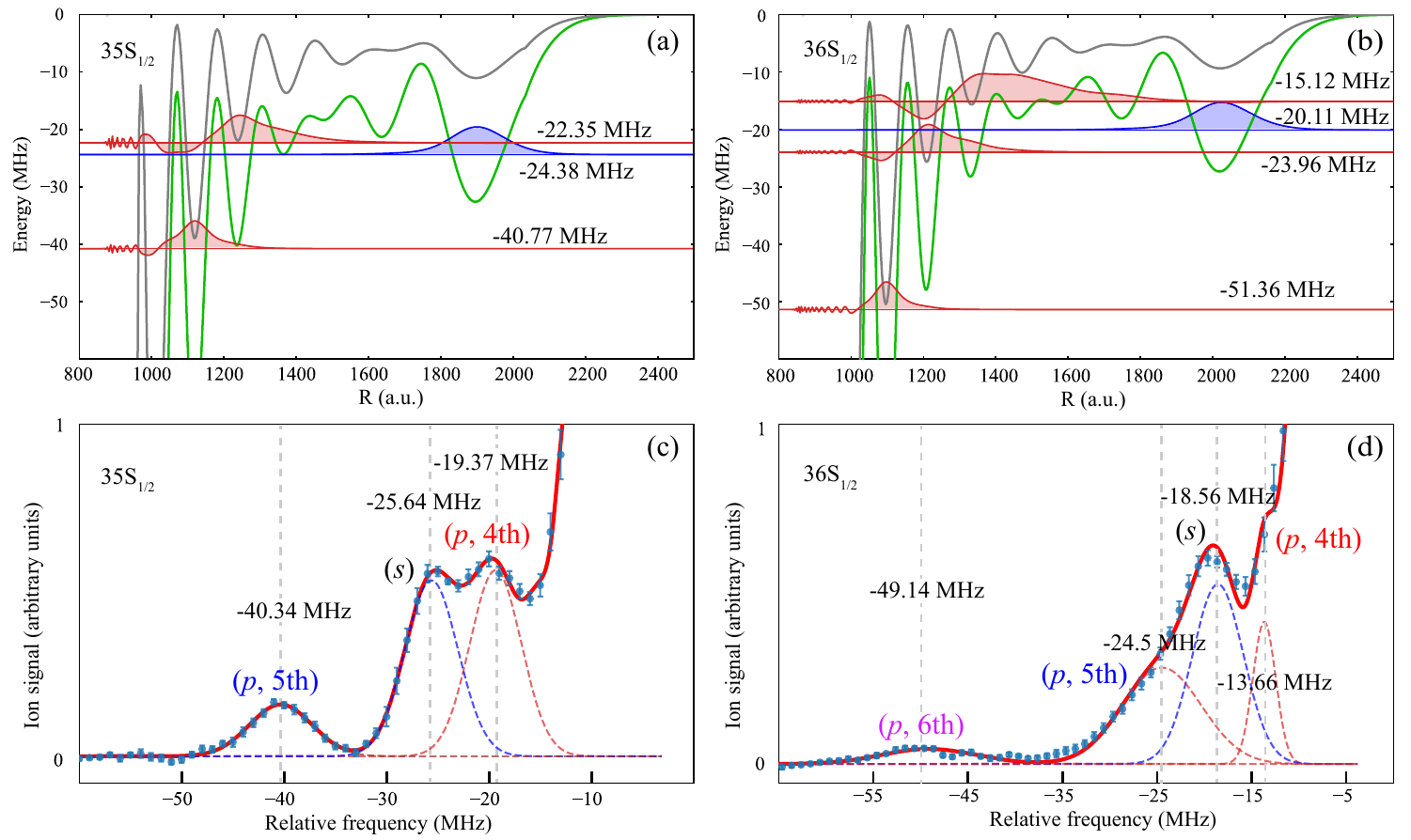}
    \caption{\textbf{Red-detuned $s$-wave shift and multi-channel $s$-$p$ coupling.}
Panels (a) and (b) show the molecular adiabatic potential-energy curves for the \(35S_{1/2}\) and \(36S_{1/2}\) states, with the measured \(s\)-wave and \(p\)-wave vibrational states observed in panels (c) and (d) indicated. The corresponding vibrational levels are labeled, and the energies are referenced to the corresponding Rydberg state. In panel (c), the 4th-node \(p\)-wave state lies above the \(s\)-wave level, so that the \(s\)-wave level is shifted towards red detuning, in accordance with the \(V_s<V_p\) case. In panel (d), the 5th- and 4th-node \(p\)-wave states lie on opposite sides of the \(s\)-wave level and are both energetically close to it, so that the \(s\)-wave level is simultaneously coupled to two \(p\)-wave states, corresponding to the multi-channel regime. The fitted peaks are assigned to \(s\)-wave states and \(i\)th \(p\)-wave states; their fitted energy positions are marked by dashed vertical lines, and the fitted energies are labeled. The data in panels (c) and (d) were obtained from 10 and 5 independent experimental runs, respectively.}
    \label{fig.3}
\end{figure*}

To describe this situation, we extend the model to a two-channel formulation in the basis \(\{|s\rangle,|p\rangle\}\):
\begin{equation}
H_{\text{sp}} =
\begin{pmatrix}
E_s & U_{\text{sp}} \\
U_{\text{sp}} & E_p
\end{pmatrix},
\label{eq:diabatic_H}
\tag{3}
\end{equation}
where $U_{\text{sp}}$ arises from the interaction between \textit{s}- and \textit{p}-wave scattering channels. These partial waves being described by good quantum numbers only for the isolated free-electron–perturber scattering problem, while in the molecule the ionic core breaks the spherical symmetry and the two channels mix automatically in the full pseudopotential Hamiltonian. Diagonalizing Eq.~\eqref{eq:diabatic_H} gives the energies
\begin{equation}
E_{\pm} = \frac{E_s + E_p}{2}
\pm \sqrt{\left(\frac{E_s - E_p}{2}\right)^2 + U_{\text{sp}}^2}.
\label{eq:adiabatic_PEC}
\tag{4}
\end{equation}
The relative magnitude of \(E_s\) and \(E_p\) determines the direction of the level shift induced by the coupling. For \(E_s-E_p>0\), the \(p\)-wave level lies below the \(s\)-wave level, giving \(\Delta E_s= E_+ - E_s>0\) and pushing the \(s\)-wave level towards blue detuning; for \(E_s-E_p<0\), the \(p\)-wave level lies above and \(\Delta E_s=E_- - E_s<0\), pushing the \(s\)-wave level towards red detuning. In both cases, the \(s\)-wave level is repelled away from the approaching \(p\)-wave level. This level repulsion manifests directly as an avoided crossing, as shown in Fig.~\ref{fig.1}(c), where the calculated \(E_s\) and \(E_p\) energies repel each other with a minimal gap of \(2|U_{\text{sp}}|\) at \(E_s=E_p\). In a single-channel picture, the two levels would cross freely; the observed avoided crossing and the resulting mixed character of the states therefore demonstrate that the \(s\)- and \(p\)-wave channels are coupled.

\begin{figure*}[t]
    \centering 
    \includegraphics[width=1\textwidth]{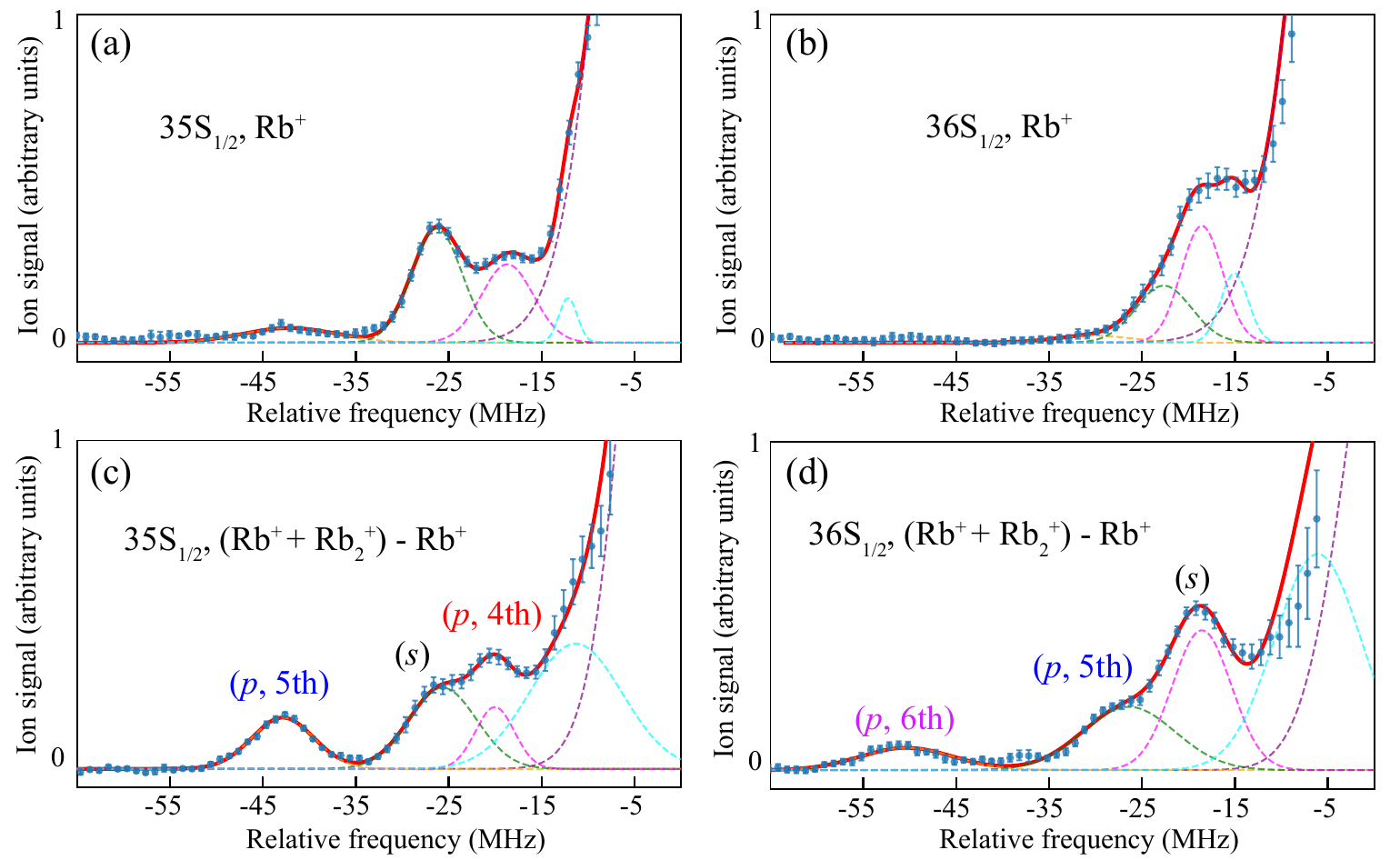}
    \caption{\textbf{Identification of coupled s- and p-wave molecules in the ion spectra of Rb\(^+\) and Rb\(_2^+\).}(a) and (b) show the ion spectra of Rb\(^+\) for the \({35}\mathrm{S}_{1/2}\) and \({36}\mathrm{S}_{1/2}\) states, respectively. (c) and (d) present the (Rb\(^+\) + Rb\(_2^+\)) - Rb\(^+\) spectra for the \(^{35}\mathrm{S}_{1/2}\) and \(^{36}\mathrm{S}_{1/2}\) states, respectively. The blue dots represent the mean values and error bars obtained from 10 independent experimental trials. The red solid lines are the results of multiple-Gaussian fitting, and the dashed lines denote the individual Gaussian fitting curves. The resonance peaks assigned to $s$-wave and $p$-wave are also labeled.}
    \label{fig.4}
\end{figure*}

\subsection*{Steeply scaling of \(p\)-wave molecules}
To investigate how the \(s\)- and \(p\)-wave molecular states evolve with the principal quantum number and exhibit the coupling, we perform high-resolution photoassociation spectroscopy on Rb(\(nS\))-Rb(\(5S\)) ultralong-range Rydberg molecules for \(n=31\) to 36. Figure~\ref{fig.2}(a) presents the radial wavefunctions of Rydberg atoms with \(n=34\) (blue) and \(n=35\) (red), respectively. The 5th wavefunction nodes (marked by blue and red circles) occur at different internuclear distances for \(n=34\) and \(n=35\), directly determining the local gradient \(|\nabla \psi_{ns}(R)|^2\) that governs the \(p\)-wave scattering contribution through the second term of the Fermi pseudopotential. These distinct nodal positions give rise to \(p\)-wave molecular states with different binding energies for the two \(n\) values.

Figure~\ref{fig.2}(b) shows the calculated molecular adiabatic potential energy curves (PECs) for \(n=34\) (blue) and \(n=35\) (red) as a function of internuclear distance \(R\), obtained from the Fermi pseudopotential. The potential wells dominated by \(s\)-wave and \(p\)-wave interactions are distinguished by their interatomic separations: the \(s\)-wave well, proportional to \(|\psi_{ns}(R)|^2\), appears at larger \(R\) and is relatively shallow; the \(p\)-wave well, proportional to \(|\nabla \psi_{ns}(R)|^2\), emerges at shorter internuclear distances where the Rydberg electron wavefunction oscillates rapidly. As a result, the \(p\)-wave potential wells for \(n=34\) and \(n=35\) are staggered at shorter internuclear distances \(R\), exhibiting a large difference in binding energies; see the shaded areas in Fig.~\ref{fig.2}(b).

We further measured Rydberg molecule spectra for \(n=31\) to 36, as shown in Fig.~\ref{fig.2}(c). The \(s\)- and \(p\)-wave molecules exhibit markedly different dependencies on \(n\): a shallow dimer state at smaller detuning—predominantly governed by \(s\)-wave scattering—appears universally across all measured \(n\) with comparable intensity and lineshape, whereas the deeply bound state at larger detuning exhibits a striking dependence on \(n\), with the \(p\)-wave molecules corresponding to the \(i\)-th node appearing staggered in the spectra. The scaling of the measured \(s\)- and \(p\)-wave molecules with \(n\) is presented in Fig.~\ref{fig.2}(d), in which the \(s\)-wave molecule and the 4th- and 5th-node \(p\)-wave molecule resonant frequencies are plotted against the principal quantum number \(n\). In this panel, the resonant peaks of the \(s\)-wave molecules show only a weak dependence on \(n\), whereas those of the 4th- and 5th-node \(p\)-wave molecules exhibit a much stronger variation with \(n\).

Both sets of observed data support a monotonic scaling of \(n\) when analyzing the same node signal. When considering the adjacent signal near the \(s\)-wave molecule signal, for example, the \(i\)-th node (\(i=3,4,5\)) \(p\)-wave molecule signals for odd \(n\), the scaling follows a function of \(n^{*-4.69}\) [gray line in Fig.~\ref{fig.2}(d)], consistent with the odd–even staggered effect noted in Ref.~\cite{engel2019precision}. The node-index dependence reveals the scaling for the \(p\)-wave case, yielding \(n^{*-14.54}\) for the 4th node, and yielding \(n^{*-17.69}\) for the 5th node. This marked difference between the \(s\)-wave and \(p\)-wave scaling exponents is the key to the \(s\)-\(p\) coupling observed here: because the two channels scale so differently with \(n\), their molecular potentials move relative to each other and can become energetically proximate at certain \(n\), where an avoided crossing between the two channels emerges. This marked difference is further corroborated by the theoretical calculations presented in the Supplementary Materials.

\subsection*{Coupling between $s$- and $p$-wave molecules}
The node-dependent behavior also provides access to the coupling between the \(s\)- and \(p\)-wave channels. As \(n\) changes, the \(s\)-wave energy \(E_s\) and the separate \(p\)-wave potential \(E_p\) move relative to one another, so that their crossing point \(R_c\) is displaced. Whenever a vibrational level bound by the \(i\)-th \(p\)-wave well comes close in energy to one bound by the \(s\)-wave well, the coupling \(U_{\text{sp}}\) admixes the two states and an avoided crossing appears in the vibrational spectrum. This mixing is especially strong for low-node \(p\)-wave molecules, since their binding energies are closer to those of the \(s\)-wave molecules; under these conditions the two channels lie energetically adjacent, the coupling is substantially amplified, and the resulting vibrational levels acquire a mixed \(s\)-\(p\) character through the avoided crossing. The \(s\)-\(p\) coupling introduces a slight discrepancy between the theoretically calculated scaling and the experimental results. 

As a direct spectroscopic consequence, the \(s\)-wave molecular level for even \(n\) is shifted towards blue detuning by this \(s\)-\(p\) coupling, see the measured spectra of Fig.~\ref{fig.2}(c). At $n=32$, the $p$-wave molecular state associated with the 3rd node shifts the $s$-wave molecular level, displacing it by about \(\Delta E_s=6.52\)$~\text{MHz}$ towards blue detuning. Similarly, at $n=34$, the $p$-wave state associated with the 4th node shifts the $s$-wave molecular level by about \(\Delta E_s=1.68\) MHz towards blue detuning. In both cases, the $s$-wave level is pushed away from the approaching $p$-wave level rather than crossing it, which is the hallmark of an avoided crossing induced by $U_{\text{sp}}$. This behavior corresponds to the \(E_s>E_p\) case discussed above: since the \(p\)-wave level lies below the \(s\)-wave level in energy, the coupling pushes the \(s\)-wave level upwards, i.e. towards blue detuning. As a result, the two states are no longer of pure character: the \(s\)-wave-dominated level acquires an admixture of \(p\)-wave character, while the \(p\)-wave-dominated level contains a finite \(s\)-wave component. Such a level repulsion and the accompanying redistribution of spectral intensity cannot be captured by considering the independent scattering channels, and instead reflect the mixed $s$-$p$ character of the resulting vibrational states. This $s$-$p$ crossover therefore provides a spectroscopic signature of the coupling between the two scattering channels.

After optimizing the experimental parameters, we tuned the principal quantum number to \(n=35\), where the situation is reversed. The \(p\)-wave molecular state associated with the 4th node now lies above the \(s\)-wave molecular level in energy, so that \(E_s<E_p\) [see the theoretical results in Fig.~\ref{fig.3}(a)]. According to the discussion above, the \(s\)-wave-dominated level is then pushed downwards, i.e. towards red detuning, away from the approaching \(p\)-wave level, as shown in Fig.~\ref{fig.3}(b). This is the opposite of the blue-detuned shift observed at \(n=32\) and \(n=34\), and it confirms that the direction of the level shift is controlled by the relative magnitude of \(E_s\) and \(E_p\), as expected for an avoided crossing induced by \(U_{\text{sp}}\). The theoretically calculated energy of the \(s\)-wave molecular level is $E_s$ = \(-24.38~\mathrm{MHz}\), whereas the experimentally measured value is \(-25.64~\mathrm{MHz}\). The experimental level lies about \(1.26~\mathrm{MHz}\) below the calculated one, i.e. shifted towards red detuning, in agreement with the \(E_s<E_p\) case. According to this experimental data, we obtained an experimental $s$-$p$ coupling strength of $U_\text{sp}$ = 2.04 MHz via Eq.~\eqref{eq:adiabatic_PEC}. This observation, together with the blue-detuned shifts at \(n=32\) and \(n=34\), demonstrates that the \(s\)-\(p\) coupling can shift the \(s\)-wave level in both directions depending on the sign of \(E_s-E_p\), providing a consistent spectroscopic signature of the avoided crossing between the two channels.

A more complex situation arises when the principal quantum number is tuned to \(n=36\). Here the 5th-node \(p\)-wave state lies below the \(s\)-wave level while the 4th-node \(p\)-wave state lies above it, so that the \(s\)-wave level is simultaneously coupled to two \(p\)-wave states on opposite sides, as shown in Fig.~\ref{fig.3}(c). In this regime the simple two-channel picture is no longer sufficient: the \(s\)-wave level is repelled from below by the 5th-node \(p\)-wave state and from above by the 4th-node \(p\)-wave state, and the resulting level shifts and mixed characters reflect the interplay of three coupled channels, see the results given in Fig.~\ref{fig.3}(d). This observation extends the \(s\)-\(p\) coupling discussed above to a multi-channel scenario, in which a single \(s\)-wave state interacts with two \(p\)-wave states of opposite detuning, and provides a natural route toward more complex coupled molecular spectra in ultralong-range Rydberg molecules.

\subsection*{Mass-selective identification}
As the \(s\)-\(p\) coupling becomes observable only when \(E_s\) and \(E_p\) are energetically close, accurately identifying the respective frequency positions of the \(s\)- and \(p\)-wave molecules is important. The weakly bound \(s\)-wave states tend to dissociate upon ionization and are detected as atomic Rb\(^+\) ions, whereas the deeply bound \(p\)-wave states retain their molecular bond and form Rb\(_2^+\) ions. In the experiment, via a mass-resolved channel-selective detection method, we can record the Rb\(^+\) and Rb\(_2^+\) ion signals. By comparing the spectral features in these two mass-selected channels, we can separate their overlapping signals and unambiguously identify the \(s\)-\(p\) avoided crossing shift, ruling out artifacts from broad linewidths.

We demonstrate the effectiveness of this method using the \(n=35\) and \(n=36\) cases as examples, where the \(s\)- and \(p\)-wave levels are expected to be strongly coupled. The results are shown in Figs.~\ref{fig.4}(a-d). Panels (a) and (b) display the Rb\(^+\) ion spectra for the \({35}S_{1/2}\) and \({36}S_{1/2}\) states, respectively, whereas panels (c) and (d) present the (Rb\(^+\) + Rb\(_2^+\)) - Rb\(^+\) spectra, which isolate the molecular contribution by subtracting the atomic background. In the total ion signal, the \(s\)- and \(p\)-wave resonances are either buried in the atomic background or severely overlap with each other. In contrast, in the mass-selected molecular spectra, these resonances become clearly distinguishable. These examples demonstrate that mass-selected ion detection provides a reliable way to resolve the coupled \(s\)- and \(p\)-wave states, which is essential for extracting their energy shifts and confirming the avoided crossing behavior.

\section*{Discussions}
In this work, the experimental conclusion is drawn primarily from the frequency shift of the \(s\)-wave molecular level. This choice is motivated by the fact that the diagonalization method employed here is accurate for calculating the \(s\)-wave molecular energy: the \(s\)-wave channel is well described by the diagonal part of the Hamiltonian, and its binding energy is only weakly affected by the detailed treatment of the \(p\)-wave channel in the absence of coupling. The \(s\)-wave level therefore provides a reliable reference against which the effect of \(s\)-\(p\) coupling can be quantified. It should be noted, however, that the \(s\)-\(p\) coupling is inherently reciprocal: just as the \(s\)-wave level is shifted by the presence of the \(p\)-wave level, the \(p\)-wave molecular level is also shifted by the coupling. The \(p\)-wave shift is more difficult to isolate experimentally, because the \(p\)-wave potential itself depends sensitively on the energy-dependent scattering length \(a_p(E)\) and on the detailed shape of the Rydberg electron wavefunction, both of which are affected by the same coupling. For this reason, we focus on the \(s\)-wave shift as the clearest spectroscopic signature of the \(s\)-\(p\) coupling, while recognizing that the \(p\)-wave states are equally subject to the level repulsion induced by \(U_{\text{sp}}\).

The \(s\)-\(p\) coupling observed here is not restricted to a single pair of channels. When several \(p\)-wave vibrational levels lie close in energy to the \(s\)-wave level, the \(s\)-wave state can couple to more than one \(p\)-wave state simultaneously, as is the case at \(n=36\), where the 4th- and 5th-node \(p\)-wave states lie on opposite sides of the \(s\)-wave level and are both energetically close to it. This multi-channel coupling modifies the level shifts and mixed characters of the resulting vibrational states in a way that cannot be captured by a simple two-level description. The present results therefore provide a natural starting point for extending the \(s\)-\(p\) coupling framework to multi-channel molecular spectra in ultralong-range Rydberg molecules. Furthermore, the mechanism demonstrated here is not limited to rubidium but can be extended to other alkali species, polyatomic molecules \cite{rittenhouse2010ultracold,rittenhouse2011ultralong,gaj2014molecular,fey2019effective,guttridge2023observation}, or heteronuclear molecules \cite{eiles2018twocomponent,wojciechowska2025ultralong,whalen2020heteronuclear,peper2021heteronuclear,guttridge2025individual,li2025diatomic}, providing a general spectroscopic approach for studying coupled molecular states across diverse atomic and molecular platforms.

\section*{Summary}
In this work, we report the experimental observation of \(s\)-\(p\) coupling in Rb(\(nS\))-Rb(\(5S\)) ultralong-range Rydberg molecules using high-resolution spectroscopy. Our measurements reveal two distinct types of molecular resonances: a shallow dimer state predominantly governed by \(s\)-wave scattering, which appears universally across all measured principal quantum numbers, and a deeply bound state associated with the \(p\)-wave scattering channel, whose position depends sensitively on \(n\). As \(n\) varies, the \(s\)- and \(p\)-wave potentials move relative to each other and can become energetically proximate, so that the off-diagonal coupling \(V_{\text{sp}}\) admixes the two channels and opens an avoided crossing in the vibrational spectrum. This avoided crossing is most pronounced for low-node \(p\)-wave molecules, whose binding energies lie closer to those of the \(s\)-wave molecules, and manifests as a blue-detuned shift of the \(s\)-wave level at \(n=32\) and \(n=34\), while the reversed situation at \(n=35\) produces a corresponding red-detuned shift. The resulting level repulsion, redistribution of spectral intensity, and mixed \(s\)-\(p\) character cannot be captured by the independent scattering channels.

\section*{Methods}
\subsection*{Experimental setup}
The cold sample of ${}^{85}\mathrm{Rb}$ atoms was produced in a MOT with a magnetic offset field $B_0 \approx 2~\mathrm{G}$, where the atoms were trapped in the ground state $5S_{1/2}$, $F=3$, $m_f=3$. At a temperature $T$ = 16 $\mu$K, the approximately spherical atomic cloud had a Gaussian radius of 300 $\mu$m. After the initial MOT loading phase of 16.4 ms, we apply a polarization gradient cooling stage of 3 ms, producing the final cold sample. During the entire preparation sequence, a crossed optical dipole trap at 1064 nm (precilasrs, FA-SF-1064-50-CW) is maintained to achieve a higher atomic density of $\rho_{G}$ = 5.2 $\times\text{ }10^{11}\text{ cm}^{-3}$ by providing additional confinement. The dipole trap has a beam waist of approximately 35 $\mu$m and a total power of 15 W, and was switched off 10 $\mu$s before the excitation pulse. 

The Rydberg state $nS_{1/2}$ was addressed by a two-photon excitation $5S_{1/2}\rightarrow5P_{3/2}\rightarrow nS_{1/2}$ using continuous-wave lasers at 780 nm (UniQuanta, ECL801-780 and TAL801-780) and 481.4 - 483.2 nm (Toptica, TA-SHG pro). The laser power of the red laser was 50 $\mu$W in a $1/e^{2}$ radius of 50 $\mu$m and the power of the blue laser was 100 mW in a $1/e^{2}$ radius of 30 $\mu$m. To avoid resonant scattering, the probe laser was blue-detuned by 220 MHz from the intermediate level $5P_{3/2}$. After a $3\text{ }\mu\text{s}$ excitation pulse and a $5\text{ }\mu\text{s}$ delay, ions originating from the Rydberg atoms or molecules were accelerated toward a microchannel plate (MCP(North Night Vision Technology, LC)) by a $5 \text{V/cm}$ electric field. The electric fields required during the experiment are established by a pair of in-vacuum electrode plates driven by a signal generator. 

 \subsection*{Automated Measurement and Control}

As described above, ions originating from the Rydberg atoms or molecules were accelerated by an extraction electric field and subsequently detected by an MCP. Each individual ion produced a single current pulse through the electron avalanche process inside the MCP. In our experiment, however, multiple ions typically arrived at the MCP within a very short time interval, resulting in an envelope-like current signal. To extract the ion signal, the integration gate of a Boxcar Averager (SRS, SR200 Series) was positioned at the maximum of the signal envelope. In this way, the MCP output could be integrated and averaged, producing a real-time voltage signal proportional to the ion yield. The voltage signal is then acquired by a data acquisition card (NI, USB-6341), transferred to the computer, and further processed and stored.

Meanwhile, the computer can also generate arbitrary waveform signals through a signal generator (Rigol, DG822 Pro and Tektronix, AFG 3252) to control the acousto-optic modulator (AOM(Gooch \& Housego, 3100-125 and A.A, MT80-B30A1-IR)). All experimental devices are synchronously triggered by a timing card (NI, PCI-6602). The AOM enables both amplitude modulation and frequency modulation of the lasers, including the probe and dipole-trap beams. In addition, the input signals of the high-voltage amplifier (Aigtek, ATA-2168) are also provided by the signal generator, allowing for flexible control of the extraction electric field and the background electric field during excitation. Within the entire experimental system, the computer acted simultaneously as both the data acquisition terminal and the signal control terminal. By coordinating these devices through custom-designed Python control programs, fully automated data acquisition and experimental control can be realized.

\subsection*{Theoretical Model and Computational Method}
Ultralong-range Rydberg molecules consist of a Rydberg atom and one or more ground-state atoms bound through low-energy electron--atom scattering interactions \cite{greene2000creation}. Taking the ionic core as the coordinate origin, $\vec{r}$ and $\vec{R}$ denote the coordinates of the Rydberg electron and the ground-state atom, respectively. The molecular Hamiltonian is
\begin{equation}
H = H_0(\vec{r}) + V_{ea}(\vec{R}-\vec{r}) + H_{HF} + H_{vib},
\tag{5}
\end{equation}
where \(H_0\), \(V_{ea}\), \(H_{\mathrm{HF}}\), and \(H_{vib}\) denote the isolated Rydberg-atom Hamiltonian, the electron--atom scattering interaction, the hyperfine interaction of the ground-state atom, and the vibrational energy, respectively. The first three terms determine the Born-Oppenheimer PECs, while the last term gives rise to the vibrational levels and corresponding vibrational spectra. In the high-\(n\) regime considered here, the ionic-core interaction, the hyperfine structure of the Rydberg atom, and the rotational energy are neglected.

The eigenenergies of alkali-metal Rydberg states are $E_{nlj} = -1/[{2(n-\delta_{nlj})^2}]$, where \(n\), \(l\), and \(j\) are the principal, orbital, and total angular momentum quantum numbers, and \(\delta_{nlj}\) is the quantum defect \cite{li2003millimeter}. The hyperfine interaction of the ground-state atom is described by
$H_{\mathrm{HF}} = a_{\mathrm{HF}}\,\vec{I}\cdot\vec{s}_2$
with corresponding hyperfine energies $E_{\mathrm{HF}} = a_{\mathrm{HF}}\left[F(F+1)-I(I+1)-s_2(s_2+1)\right]/2$, where \(F=|\vec{I}+\vec{s}_2|\). In our experiment, we only study the pure triplet states because of the binding energy of the mixed singlet triplet states is near to resonance.

Following Fermi and Omont \cite{fermi1934sopra,omont1977theory}, the electron--atom interaction is modeled using a zero-range pseudopotential including both \(s\)-wave and \(p\)-wave scattering contributions \cite{greene2000creation,hamilton2002shape}. To incorporate the dependence on the total angular momentum \(J\) of the scattering channels \cite{khuskivadze2002adiabatic}, we adopt the formulation developed by Eiles \textit{et al.} \cite{eiles2017hamiltonian},
\begin{equation}
\begin{aligned}
V_p
=
\delta(\vec{R}-\vec{r})
\sum_{LSJ\Omega}
\sum_{M_L,M_L'}
\mathcal{A}(SLJ,k)
C_{LM_L,S(\Omega-M_L)}^{J\Omega}
\\
\times
C_{LM_L',S(\Omega-M_L')}^{J\Omega}
\overset{\leftarrow}{\nabla}{}^{(L)}_{M_L}
|S(\Omega-M_L)\rangle
\overset{\rightarrow}{\nabla}{}^{(L)}_{M_L'}
\langle S(\Omega-M_L')|,
\label{eq:Vp}
\end{aligned}
\tag{6}
\end{equation}
where $\mathcal{A}(SLJ,k)=(2L+1)\,2\pi\,a(SLJ,k)$, and \(a(SLJ,k)\) denotes the energy-dependent scattering length or scattering volume. The relevant scattering channels are \(^{1}S_0\), \(^{3}S_1\) for \(L=0\) and \(^{1}P_1\), \(^{3}P_{0,1,2}\) for \(L=1\).

Under the Born--Oppenheimer approximation, the nuclear vibrational motion is neglected, reducing the problem to the electronic Hamiltonian at fixed \(R\). We employ the basis  $|nljm_j\rangle \otimes|m_{s_2},m_{i_2}\rangle$, where \(|nljm_j\rangle\) are the fine-structure eigenstates of the Rydberg atom, and \(|m_{s_2},m_{i_2}\rangle\) describe the electron and nuclear spins of the ground-state atom. Since only \(L=0\) and \(L=1\) scattering channels are included, it is sufficient to retain basis states with \(|m_j|\leq 3/2\).

The Rydberg wave functions are expressed as $\Psi_{nljm_j}(\vec{r})=\sum_{m_l,m_s}C_{lm_l,sm_s}^{jm_j}R_{nlj}(r)Y_{lm_l}(\vec{r})|m_s\rangle$, where \(C_{lm_l,sm_s}^{jm_j}\) are Clebsch--Gordan (CG) coefficients, \(R_{nlj}(r)\) is the radial wave function, and \(Y_{lm_l}(\vec{r})\) is the spherical harmonic. The radial wave functions are obtained using the ARC package \cite{vsibalic2017arc}. In this basis, \(H_0\) is diagonal, while the hyperfine interaction is evaluated through CG coupling coefficients. The matrix elements of \(V_p\) are calculated following Eiles \textit{et al.} \cite{eiles2017hamiltonian} using
\begin{equation}
Q^{nlj}_{LM_L}(R)
=
\delta_{m,M_L}
\left[
\vec{\nabla}^{\,L}
\phi_{nljm}(\vec{R})
\right]^L_{M_L},
\tag{7}
\end{equation}
with $\phi_{nljm}(\vec{R}) = R_{nlj}(R)Y_{lm}(\vec{R})$. The total Hamiltonian matrix is therefore written as $H = H_0 + H_{HF} + V_p$. The molecular Hamiltonian is diagonalized at each internuclear distance \(R\), thus we can obtain the PECs.

The nuclear vibrational motion is then treated within the Born--Oppenheimer approximation by solving the vibrational Schrödinger equation
\[
\left[-\frac{\nabla_R^2}{2M}+U(R)\right]\chi_v(R)=E_v\chi_v(R),
\tag{8}
\]
where \(U(R)\) denotes the calculated PECs, while \(E_v\) and \(\chi_v(R)\) are the vibrational energy and wave function for vibrational quantum number \(v\), respectively. Here, $M=(m_1m_2)/(m_1+m_2)$ is the reduced mass of the Rydberg and ground-state atoms. The vibrational Schrödinger equation is solved numerically using the finite difference method (FDM), which transforms the differential equation into a matrix eigenvalue problem. 

Due to the diagonalization method, the theoretical results with low principal quantum numbers deviate more from the experimental results. In our calculations, we adjust the scattering phase shift, the basis-set truncation range, the number and range of grid points in $R$ to precisely match the theoretical vibrational levels and the experimental data. 

\section*{Acknowledgements}
We thank Prof. Matthew T. Eiles from Purdue University for his constructive comments and helpful suggestions. We thank the fruitful discussions with Prof. Wei-Bin Li from University of Nottingham and Prof. Yi-Quan Zou from Beijing Quantum Institute. We acknowledge funding from the National Natural Science Foundation of China (Grant Nos. T2495253, 62435018), the National Key R and D Program of China (Grant No. 2022YFA1404002).

\hspace*{\fill}

\section*{Data Availability}
All experimental data used in this study are available from the corresponding author upon request.

\section*{Author contributions statement}
D.-S.D. conceived the idea and supported this research. Q.L. and S.Y.S. conducted the physical experiments. Q.L conducted the theoretical calculations. The manuscript was written by D.-S.D., Q.L. and S.Y.S.. All authors contributed to discussions regarding the results and the analysis contained in the manuscript.

\section*{Competing interests}
The authors declare no competing interests.

\bibliography{ref}

\end{document}


\title{Supplementary Material: Observation of $s$-$p$ Coupling in Ultralong-Range Rydberg Molecules}

\author{Qing Li$^{1,2}$}
\thanks{These authors contributed equally to this work.}
\author{Shi-Yao Shao$^{1,2}$}
\thanks{These authors contributed equally to this work.}
\author{Jun Zhang$^{1,2,3}$}
\author{Han-Chao Chen$^{1,2,3}$}
\author{Li-Hua Zhang$^{1,2,4}$}
\author{Bang Liu$^{1,2,4}$}
\author{Guang-Can Guo$^{1,2,3}$}
\author{Dong-Sheng Ding$^{1,2,3,4,\textcolor{blue}{\dagger}}$}
\author{Bao-Sen Shi$^{1,2,3}$}

\affiliation{$^1$Laboratory of Quantum Information, University of Science and Technology of China, Hefei, Anhui 230026, China.}
\affiliation{$^2$Anhui Province Key Laboratory of Quantum Network, University of Science and Technology of China, Hefei 230026, China.}
\affiliation{$^3$Hefei National Laboratory, University of Science and Technology of China, Hefei 230088, China}
\affiliation{$^4$Davinci Quantum Technology Co., Ltd, Hefei, Anhui 230026, China}

\date{\today}
\symbolfootnote[2]{dds@ustc.edu.cn}

\maketitle

This supplementary material provides the detailed information about the experimental time sequence, the experimental setup, the theoretical calculations, and the measured stark shift.

\section*{Excitation and detection}
The two-photon excitation to Rydberg states follows a ladder scheme: $|5\text{S}_{1/2},\ f=3,\ m_f=3\rangle\xrightarrow{\rm{probe}} |5\text{P}_{3/2},\ f=4,\ m_f=4\rangle\xrightarrow{\rm{coupling}}|n\text{S}_{1/2}\rangle$, as shown in Fig.~\ref{figTime}(a). The probe laser at 780 nm is blue-detuned by 220 MHz from the intermediate transition, has a power of 50$\mu$W, and is focused to a beam waist of 50$\mu$m at the atoms. Its circular polarization is $\sigma^+$ . The coupling laser at 481.4 - 483.2 nm is $\sigma^{-}$ polarized, with a power of 100 mW and a beam waist of 30 $\mu$m. The probe and coupling beams counter-propagate along the quantization axis. The probe laser is locked using saturation absorption spectroscopy, while the coupling laser frequency is scanned across the Rydberg resonance. To control the two-photon excitation, the probe excitation field is switched by an AOM, while the coupling field is gated by a combination of an EOM and a mechanical chopper.

Following the excitation pulse, a clearing electric field of 5 V/cm is applied across the electrodes to guide ions toward the MCP. The MCP signal is amplified by a transimpedance amplifier and integrated using a gated integrator for data acquisition. Rydberg molecules are then identified by scanning the coupling laser frequency. In the resulting ion spectrum, the dominant peak corresponds to the unperturbed Rydberg atomic resonance. The molecular signals manifest as additional, weaker peaks at specific red detunings relative to this atomic resonance, which can be checked by the vibrational levels obtained from theoretical calculations.

\section*{Experimental time sequence}

\begin{figure*}[htbp]
    \centering 
    \includegraphics[width=1\textwidth]{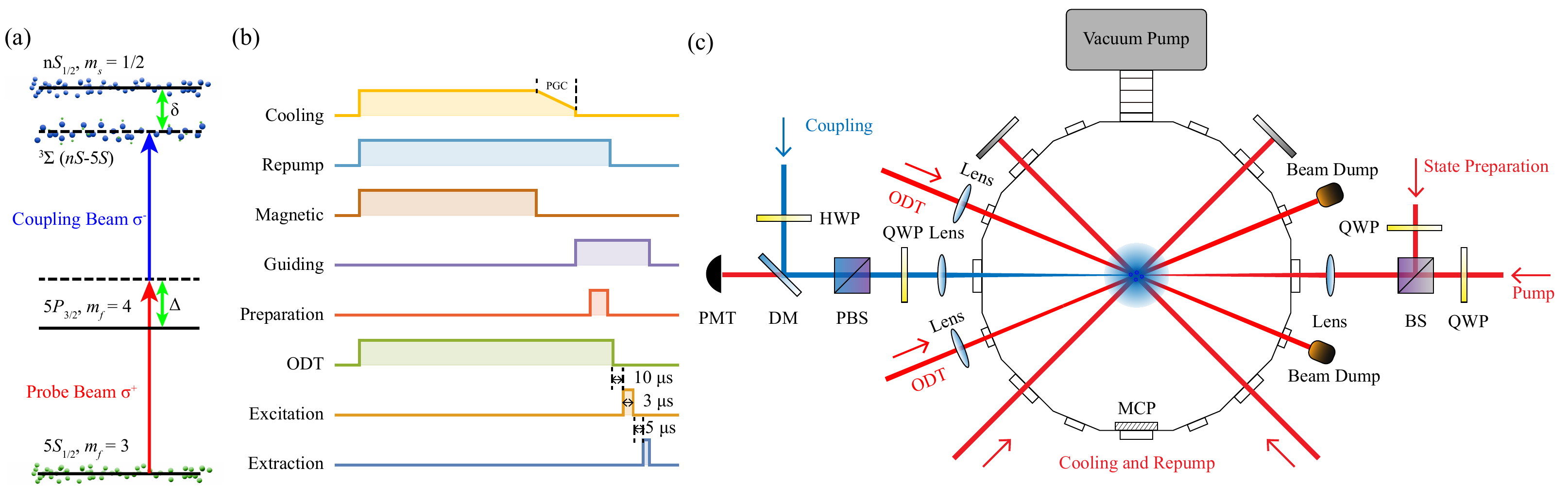}
    \caption{
    \textbf{Experimental apparatus and timing sequence} (a) Excitation scheme of the experiment. The atoms are excited from the $5S_{1/2},\,m_f=3$ ground state to the $nS_{1/2},\,m_s=1/2$ Rydberg state via a two-photon excitation process using probe and coupling laser beams. To suppress spectral broadening caused by intermediate-state scattering, the probe laser is detuned by $\Delta=+220~\mathrm{MHz}$. Molecules are formed when the two-photon detuning $\delta$ is resonant with the molecular vibrational levels. (b) Pulse sequence of the experimental protocol, detailing the chronological stages of cooling, repump, magnetic field, guiding, state preparation, optical dipole trap (ODT), excitation, and extraction. The polarization gradient cooling (PGC) process is showcased as a specific sub-phase within the cooling light sequence. Horizontal lines indicate the operational states (on/off) of all laser beams, magnetic fields, and detection components along a relative timeline, with key delays (e.g., $10\ \mu\text{s}$, $3\ \mu\text{s}$, and $5\ \mu\text{s}$) explicitly marked. (c) Schematic diagram of the experimental setup. The layout illustrates the vacuum chamber, the cooling and repump beams, the ODT beams, the coupling and pump beam paths, as well as the detection systems including the photomultiplier tube (PMT) and microchannel plate (MCP).}
    \label{figTime}
\end{figure*}

The entire experimental cycle lasts for 20 ms, corresponding to a repetition rate of 50 Hz, as shown in Fig.~\ref{figTime}(b). The experiment begins with the trapping of background $^{85}\text{Rb}$ atoms by a three-dimensional magnetic optical trap (MOT) \cite{raab1987trapping,phillips1998nobel}. The MOT remains continuously active for the initial 16.4 ms to complete the preliminary loading and precooling of the atoms.  At 16.4 ms, the MOT magnetic field is rapidly extinguished, initiating a 3 ms polarization gradient cooling (PGC) process \cite{lett1988observation,dalibard1989laser,salomon1990laser}. At 19.4 ms, the cooling light is turned off to end the PGC stage, while the repump light continues to participate in the subsequent initial state preparation process until 19.8 ms. 

To establish a quantization axis, the state preparation magnetic field coils are energized immediately after the PGC process ends at 19.4 ms and remain powered until the end of the cycle. Owing to the inductive charging effect of the coils, this magnetic field is established gradually. Once the magnetic field stabilizes, the initial state preparation light is switched on at 19.7 ms with a pulse width of 50 $\mu$s, completing the initial state preparation in cooperation with the repump light \cite{kastler1950quelques,happer1972optical}. A crossed optical dipole trap (ODT) remains active throughout the atom-preparation stage and is turned off 10 $\mu$s prior to excitation \cite{adams1995evaporative,grimm2000optical}.

The two-photon Rydberg excitation pulses are applied at 19.87 ms for a duration of 3 $\mu$s \cite{lukin2001dipole,singer2004suppression}. Following a 5 $\mu$s interval after the two-photon pulse ends, a pulsed clearing electric field is switched on and maintained until the end of the cycle. This electric field accelerates the ionization products of the Rydberg atoms and molecules toward a microchannel plate (MCP), thereby achieving the extraction and detection of the ion signals \cite{gallagher1994rydberg}.

\section*{Vacuum system and MOT}

The experiments are performed in an ultra high vacuum (UHV) metal chamber maintained at a base pressure of $2\times10 ^{-7}$ Pa by an ion pump and a getter pump. A rubidium reservoir consisting of a copper tube loaded with natural rubidium metal is connected to the chamber through a manually operated valve. Opening the valve raises the background vapor pressure for atom loading. All viewport windows are anti-reflection coated for the relevant wavelengths.

A pair of parallel copper electrodes is placed symmetrically on opposite sides of the atom trapping region. Each plate measured 118 mm $\times$ 24 mm and the plates were separated by 14 mm. Each plate featured a central racetrack-shaped aperture (57 mm $\times$ 13 mm, with 6.5 mm radius semicircles) to avoid blocking the trapping light, as shown in Fig.$\ $1(b). These electrodes are used exclusively for applying a pulsed electric field after Rydberg excitation to extract ions. A MCP detector is positioned at a distance of 103.7 mm from the trap center for ion detection.

A three-dimensional MOT for $^{85}$Rb is loaded from the background vapor. The MOT cooling light is tuned to the $|5\text{S}_{1/2},\ f=3\rangle\rightarrow |5\text{P}_{3/2},\ f=4\rangle$ cycling transition near 780 nm, with a red detuning of 12 MHz and a beam diameter of 15 mm. Repump light is resonant with the $|5\text{S}_{1/2},\ f=2\rangle\rightarrow |5\text{P}_{3/2},\ f=3\rangle$ transition. The six-beam configuration is formed using three independent incident beams, each with a power of 2 mW, which
are subsequently retro-reflected. The magnetic field gradient of the MOT is $\sim$ 16.8 G/cm, and three additional pairs of Helmholtz coils are used to compensate for residual magnetic fields. Absorption imaging of the atom cloud shows a Gaussian radius of $\sim 300\ \mu\text{m}$.

\section*{Atomic preparation}
After the MOT magnetic field is switched off, PGC is applied for 3 ms. During PGC, the cooling laser frequency is swept linearly from -12 MHz to -32 MHz relative to the cooling transition by applying a sawtooth modulation to an acousto-optic modulator (AOM). Simultaneously, the power is reduced to 80\% of the MOT loading level via amplitude modulation. The resulting cloud has a temperature of 17 $\mu$K and an average density of $2.6\ \times\ 10^{10}\ \text{cm}^{-3}$ , measured by time-of-flight imaging.

To further increase the atomic density, a crossed ODT remains continuously on throughout the MOT loading phase. The ODT is formed by two beams intersecting at an angle of $45^{\circ}$ , with a total power of 15 W and a beam waist of 35 $\mu$m at the trap center, as shown in Fig.~\ref{figTime}(c). Thus, we can obtain a higher density of $5.2\ \times\ 10^{11}\ \text{cm}^{-3}$.

Optical pumping prepares the atoms in the desired initial quantum state. A dedicated optical pumping beam pumps the atoms into the $|5\text{S}_{1/2},\ f=3,\ m_f=3\rangle$ state. A bias magnetic field defining the quantization axis is generated by a pair of Helmholtz coils. The field direction is parallel to the wavevector of the probe beam.

\section*{Vibrational levels calculation of molecular bound states}

\begin{figure*}[htbp]
    \centering 
    \includegraphics[width=1\textwidth]{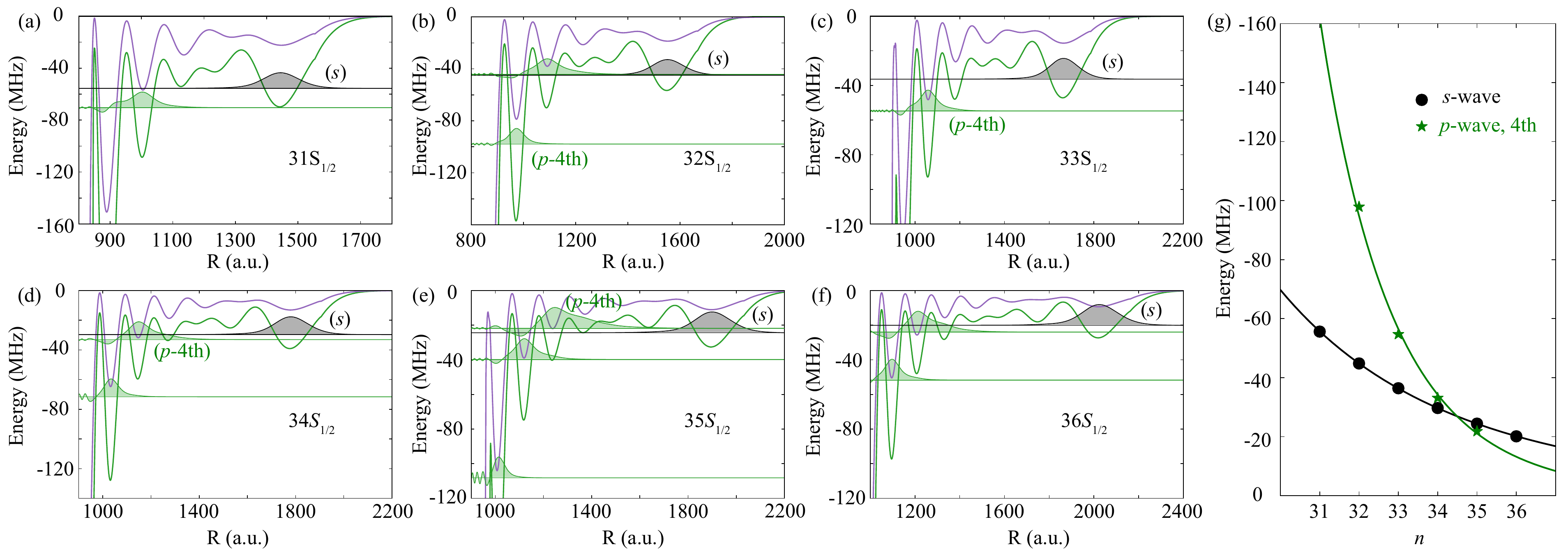}
    \caption{
    \textbf{Adiabatic potential energy curves, vibrational states, and scaling behavior of ${}^3\Sigma(5s\text{-}ns)$ Rydberg molecular states.}
    (a)-(f) The adiabatic potential energy curves of the ${}^3\Sigma(5s\text{-}ns)(\nu=0)$ molecular states with principal quantum numbers ranging from $n=31$ to $36$ are presented. The purple curves denote the singlet--triplet mixed states, while the green curves represent the pure triplet states. The $s$-wave vibrational levels and their corresponding wave functions are indicated by black and gray curves, respectively, whereas the $p$-wave vibrational levels and wave functions are shown in green. The potential wells formed by the 4th $p$-wave vibrational states are highlighted for $n$ = 32--35. (g) The scaling behavior of the fourth vibrational levels for both the $s$-wave and $p$-wave states is presented, exhibiting power-law dependences of $n^{*-6.16}$ and $n^{*-15.21}$, respectively. The calculations presented here differ slightly from those in the main text due to the use of different numerical parameters, including the basis-set truncation range and the number and spatial range of grid points along the internuclear coordinate $R$.}
    \label{theory_31-36}
\end{figure*}

\begin{table*}[htbp]
    \centering
    \caption{
    \textbf{Quantitative comparison of measured and calculated vibrational energy levels of ultralong-range Rydberg molecules.}
    Comparison between the experimentally measured and theoretically calculated vibrational energy levels of ultralong-range Rydberg molecules for principal quantum numbers $n=31$--$36$. The energies are given in units of MHz relative to the corresponding Rydberg atomic asymptotes. The $s$-wave dominated states and the $p$-wave dominated states are presented separately. For the $p$-wave states, different rows correspond to different vibrational levels, denoted as 3rd, 4th, 5th, and 6th states. The experimental values are compared with the theoretical predictions obtained from the calculated molecular potential energy curves, demonstrating the agreement between theory and experiment across different principal quantum numbers.}
    \includegraphics[width=0.8\linewidth]{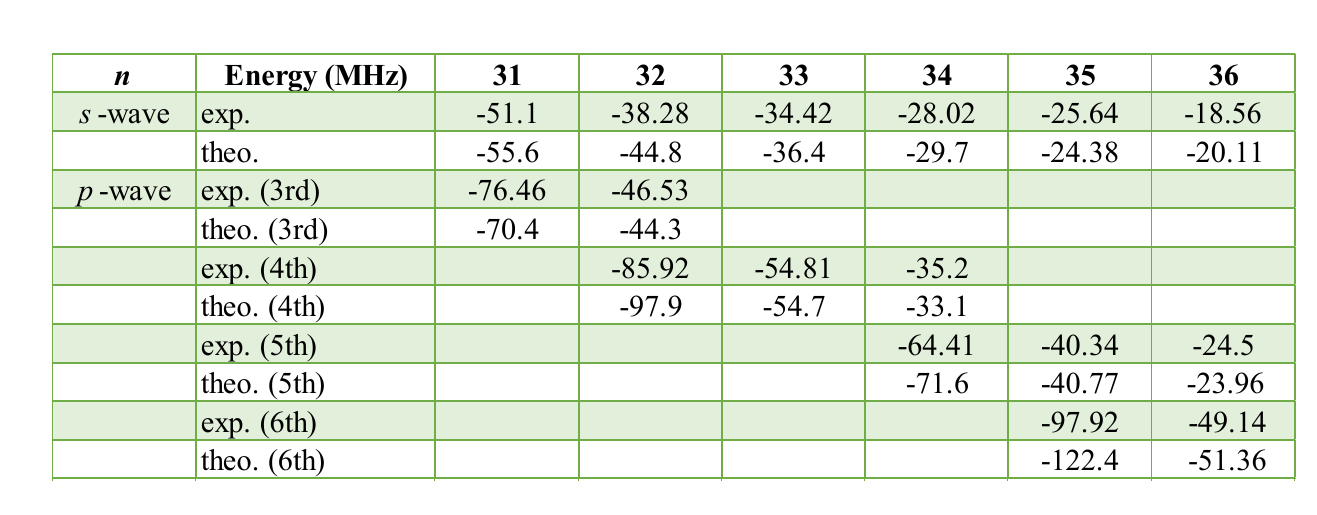}
    \label{theory_table}
\end{table*}

The adiabatic potential energy curves and vibrational levels for principal quantum numbers $n=31$ - $36$ are calculated using the diagonalization method\cite{eiles2017hamiltonian}, as presented in Fig.~\ref{theory_31-36}(a)-(f). In these calculations, the basis set is truncated within the finite range from $(n^*-1)$ to $(n^*+2.05)$, and the same set of electron-atom scattering phase shifts is employed for all cases. The number of grid points along the internuclear coordinate $R$ is fixed at 800, while the range of $R$ varies for different principal quantum numbers due to the different spatial extensions of the corresponding potential energy curves. 

The scaling behavior of the $s$-wave and fourth vibrational levels for the $p$-wave states is presented, exhibiting power-law dependences of $n^{*-6.16}$ and $n^{*-15.21}$, respectively. The numerical values of the theoretical vibrational levels obtained from Fig.~\ref{theory_31-36} and the corresponding experimental measurements are summarized in Table~\ref{theory_table}. 

The discrepancies between the theoretical predictions and experimental results mainly originate from the following factors: (1) the electron-atom scattering phase shifts have a significant influence on the calculated adiabatic potential energy curves, the scattering phase shifts used in our calculations are not perfectly precise; (2) the truncation range of the basis set can affect the accuracy of the calculated molecular states; and (3) for the $p$-wave states, the presence of a scattering resonance makes the results more sensitive to the density of the $R$-grid points.

In principle, the Green's function methods enable a more accurate calculation of the potential energy curve in the absence of an external field\cite{2023Green}. However, our experiment is demonstrated under a strong magnetic offset field $B_0 \approx 2~\mathrm{G}$ (it is difficult to calculate the binding energy via the Green's function methods),  while the diagonalization method can predict our measurements.

\begin{figure*}[t]
    \centering 
    \includegraphics[width=1\textwidth]{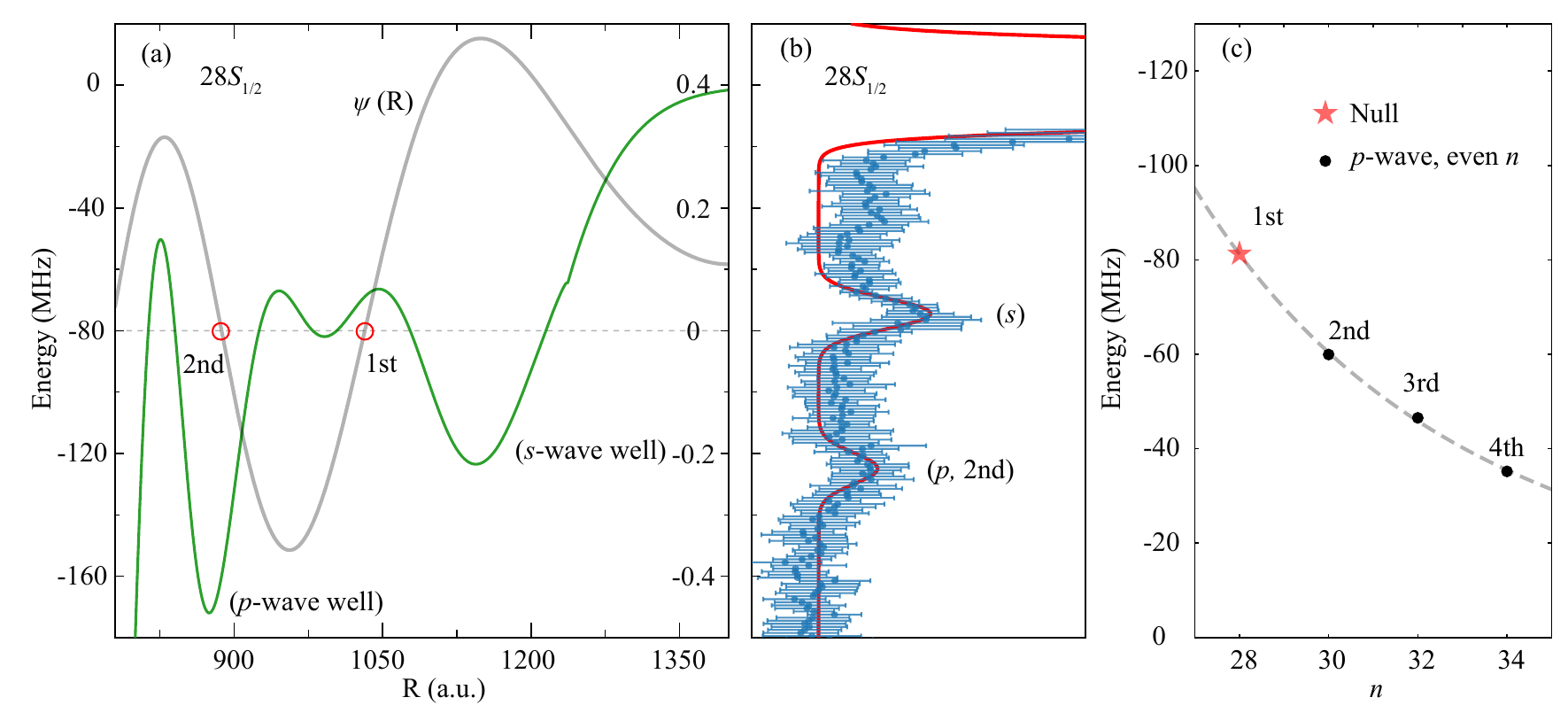}
    \caption{\textbf{Nodal-induced lower limit for the 1st-node $p$-wave molecule.} (a) The Rydberg atomic wavefunction $\psi(R)$ (gray) for 28S and calculated adiabatic potential energy curve of triplet state (green). (b) Measured molecular vibrational spectrum of the ${}^3\Sigma(28s\text{-}5s)$ bound state. The red line shows the result of Gaussian fitting. Error bars represent statistical uncertainties derived from 8 independent measurements. The spectrum includes two visible peaks corresponding to $s$-wave molecule (s) and 2nd-node $p$-wave molecule ($p$, 2nd). (c) Fitting curves for the $i$th nodes with even principal quantum numbers from 28 to 34). The null value of 1st-node $p$-wave molecule signal at $n$ = 28 confirms the theoretical prediction.}
    \label{fig.3}
\end{figure*}
\subsection*{Node-dependent $p$-wave molecule}
The deeply $p$-wave molecular states can be traced to the nodal structure of the $nS$ radial wavefunction in the inner region. In this region, the wavefunction can be approximated as $\psi_{ns}(R) \simeq \mathcal{A}_n \sin(k_n R + \delta)$, where $k_n = 1/(n^* a_0)$ and $ \mathcal{A}_n$ is the amplitude, where $n^*$ is the effective principal quantum number. By inserting this wavefunction into the gradient term of the Fermi pseudopotential, the $p$-wave contribution then becomes
\begin{equation}
V_p(R) \simeq \frac{6\pi\hbar^2}{\mu} a_p^3(E) \, \mathcal{A}_n^2 k_n^2 \cos^2(k_n R + \delta),
\label{eq:p_potential}
\tag{1}
\end{equation}
as \textit{n} increases, the number of nodes changes, and the phase of the oscillating wavefunction $\phi=k_n R + \delta$ undergoes a significant change. $V_p(R)$ is maximized at the wavefunction nodes, where $\cos^2=1$, this phase evolution directly modulates the $p$-wave potential landscape. 

The phase $\phi$ directly determines whether, at a given internuclear distance $R$, the Rydberg electron wavefunction lies near a node or near an antinode. As the principal quantum number \textit{n} increases, the radial nodes of the wavefunction continuously shift outward. When a node crosses the internuclear distance $R_{\rm{deep}}$ that supports a deep potential well (when $V_p(R) < V_s(R)$), the local gradient of the wavefunction reaches a maximum at that location, yielding a strong attractive $p$-wave contribution and a corresponding bound state at the characteristic spectral position. As \textit{n} continues to increase and the node moves away from $R_{\rm{deep}}$, the gradient at this fixed internuclear distance diminishes, and the $p$-wave contribution weakens accordingly. Coupled with the energy-dependent $p$-wave scattering length $a_p(E)$, this mechanism gives rise to internuclear-distance-dependent $p$-wave potential wells, thereby offering a sensitive and precise probe of the nodal displacement of the wavefunction. This results in the node-dependent formation of $p$-wave bound states at a well-defined internuclear distance $R$ (the distance at which molecules are effectively formed).

\subsection*{Nodal-induced lower limit}
For small principal quantum numbers $n$, the radial nodes are located at shorter internuclear distances, where the Rydberg electron's local kinetic energy—and thus the effective collision energy—pushes the scattering length towards the resonant-valued regime of a deep potential well. This makes it more challenging to observe high-node $p$-wave molecules with small $n$ under a limited atomic density. Due to the reduced number of nodes and the coarser spatial oscillation of the wavefunction at low $n$, the low-nodes appear at relatively shorter internuclear distances, resulting in a larger $a_p(E)$. As a result, the low-node $p$-wave well is sufficiently deep to bind molecular states. This is evidenced by the clearly resolved resonances associated with the 3rd, 4th, and 5th nodal  $p$-wave molecules for $n$ = 31, 33, and 35, respectively, as shown in Fig.~2(c) in main text. 

In addition, the \textit{s}-wave and the lower-node \textit{p}-wave wells become closer in distance, thereby facilitating stronger mixing states between them \cite{bendkowsky2010rydberg}. This results in the suppression of the 1st-node $p$-wave molecule for $n$ = 28 [see Fig.~\ref{fig.3}(a)]. In our extended measurements down to \textit{n} = 28, we indeed find that while \textit{p}-wave resonances remain clearly visible for the 2nd-node, they are completely absent from the spectrum for the 1st-node case, as shown in Fig.~\ref{fig.3}(b). In contrast, the shallow \textit{s}-wave dimer state persists and remains unaffected. This is illustrated in Fig.~\ref{fig.3}(c), where the absence of the characteristic 1st-node $p$-wave molecule peak for \textit{n} = 28 confirms the theoretical prediction of a nodal-induced lower limit.

\begin{figure*}[t]
    \centering 
    \includegraphics[width=1\textwidth]{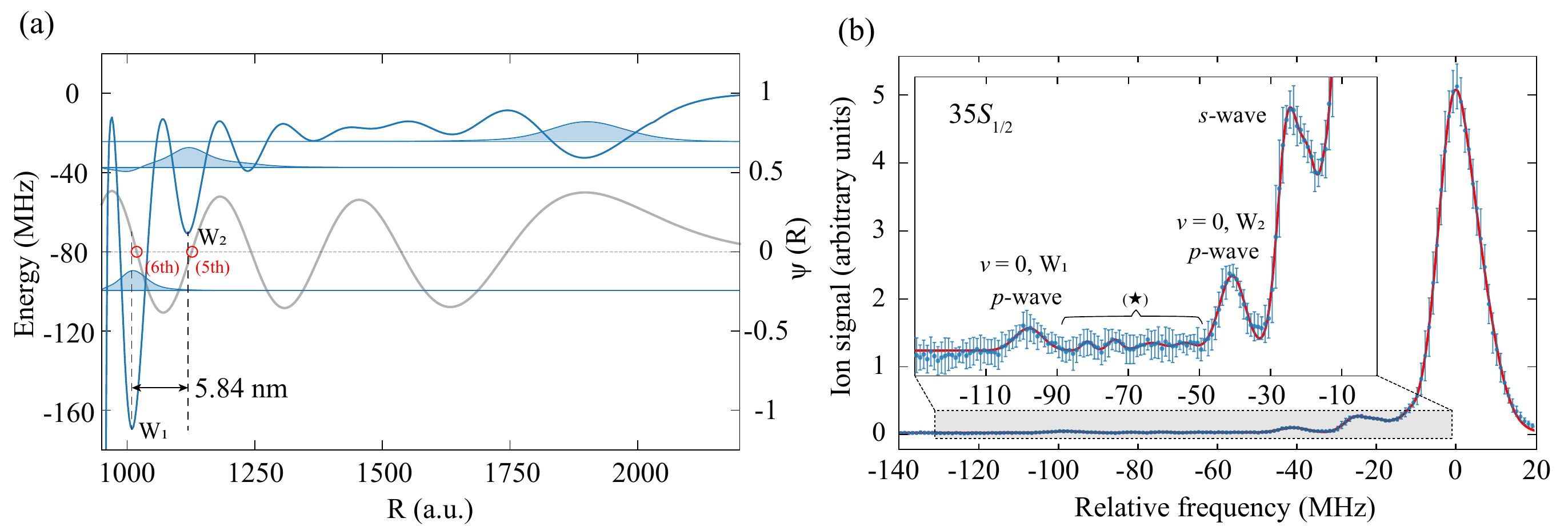}
    \caption{
    \textbf{Measurement of adjacent nodes $p$-wave molecules ${}^3\Sigma(35s\text{-}5s)$.} (a) The blue curve represents the triplet molecular potential energy curve, while the gray curve denotes the Rydberg atomic wavefunction $\psi(R)$, with the fifth and sixth nodes highlighted by red circles. Three vibrational levels are shown with energies of $-99.5~\mathrm{MHz}$, $-37.4~\mathrm{MHz}$, and $-24.3~\mathrm{MHz}$, respectively. The outermost potential well is dominated by the $s$-wave interaction, while the two inner wells are dominated by the $p$-wave interaction, labeled as W$_1$ and W$_2$. The separation between W$_1$ and W$_2$ is $5.84~\mathrm{nm}$. (b) Experimental vibrational spectra. The reference energy is set at the corresponding Rydberg atomic level. The inset shows an enlarged view of the shaded region in the main panel. The peak positions corresponding to $(p\text{-wave},\, v=0,\, \mathrm{W}_1)$, $(p\text{-wave},\, v=0,\, \mathrm{W}_2)$, and $(s\text{-wave})$ are located at $-97.92~\mathrm{MHz}$, $-40.75~\mathrm{MHz}$, and $-25.18~\mathrm{MHz}$, respectively. A series of additional unassigned peaks appearing in the range from $-50~\mathrm{MHz}$ to $-90~\mathrm{MHz}$ are marked by pentagram symbols. The red curve represents multi-Gaussian fits to the averaged data. Error bars represent statistical uncertainties derived from 8 independent measurements.
    }
    \label{fig.4}
\end{figure*}

\begin{figure*}[t]
    \centering 
    \includegraphics[width=1\textwidth]{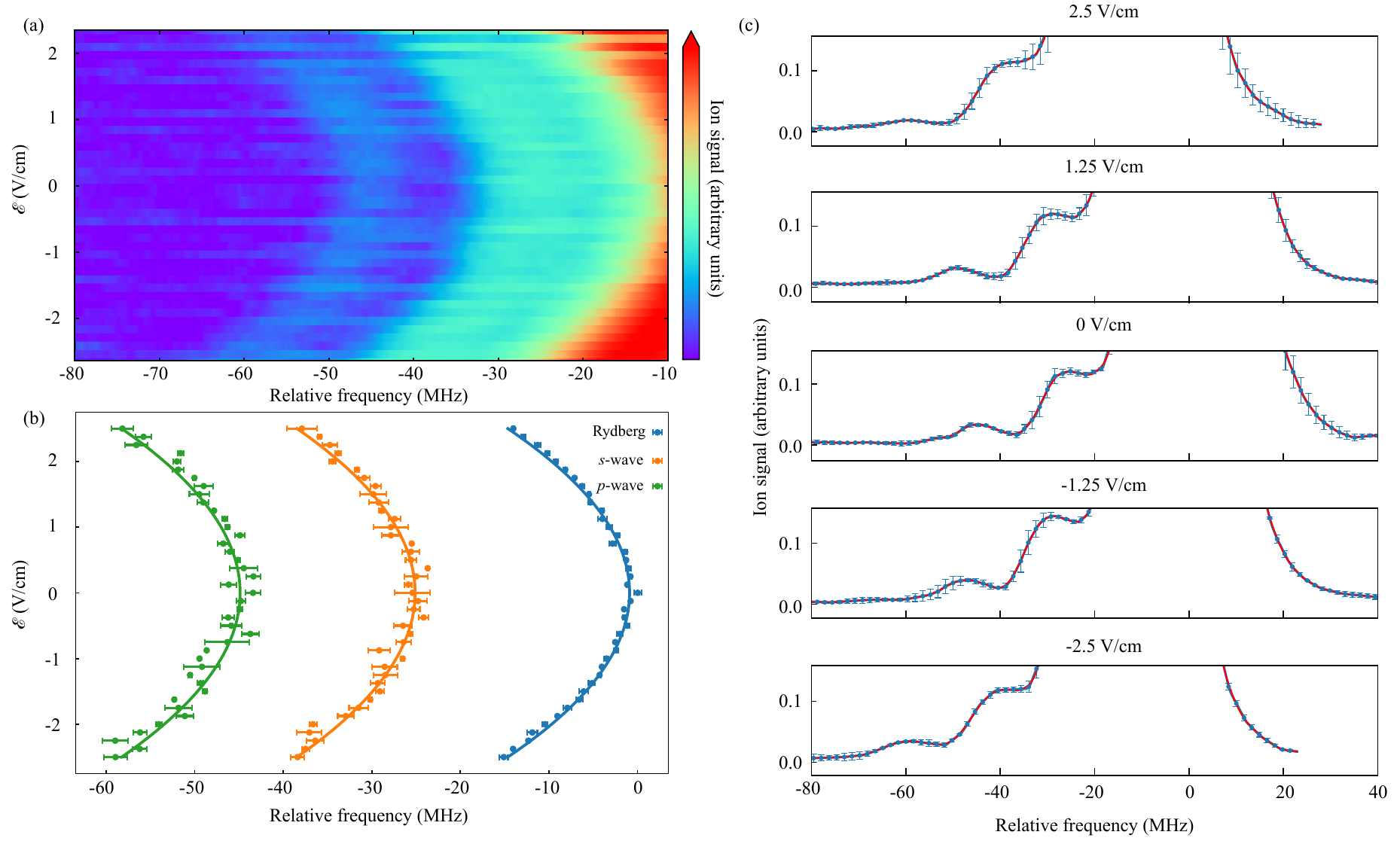}
    \caption{
    \textbf{Quadratic stark shift of molecular ${}^3\Sigma(5s\text{-}35s)(\nu=0)$ spectra in static electric fields.} (a) Molecular spectra measured under different static electric fields. The spectral features located on the left and right sides correspond predominantly to the $p$-wave and $s$-wave dominated molecular states, respectively. A quadratic frequency shift with increasing electric-field strength is observed. (b) Quadratic fitting of the spectral shifts as functions of the electric-field strength extracted from panel (a). The fitting curves follow $f = a\mathcal{E}^2 + c$, with parameters: Rydberg atomic 35s state ($a = -2.19997$, $c = -0.89532$), $s$-wave ${}^3\Sigma(5s\text{-}35s)(\nu=0)$ molecule ($a = -2.13437$, $c = -25.106669$), and $p$-wave ${}^3\Sigma(5s\text{-}35s)(\nu=0)$ molecule ($a = -2.14181$, $c = -44.86262$). (c) Representative molecular spectra at several selected electric-field strengths for $n=$ 35. The red curves represent multi-Gaussian fits to the averaged data. Error bars represent statistical uncertainties derived from 10 independent measurements.}
    \label{fig.5}
\end{figure*}

\subsection*{$p$-wave molecules at adjacent nodes}
Essentially, the observed node-dependence originates from the spatial characteristics of the $p$-wave molecular bound states. To further investigate this, we examine the $p$-wave states supported by two adjacent nodes of the Rydberg electron wavefunction.  Figure \ref{fig.4}(a) shows the radial wavefunction $\psi_{35s}(R)$ of the Rydberg electron (gray curve) and the corresponding adiabatic PEC (blue curve) for rubidium 85 in the internuclear distance range $R$ = 900 – 1300 a.u.. Within this range, two adjacent nodes of $\psi_{35s}(R)$ give rise to two distinct potential wells, labeled W$_1$ and W$_2$, with a spatial phase shift of  $\pi$. These wells arise from the $p$-wave scattering term proportional to $|\nabla \psi_{35s}(R)|^2$, which peaks near each node where the wavefunction gradient is maximal. 

Despite the large gradient magnitude at both nodes, the two wells exhibit markedly different binding energies, which arise from the energy dependence of the $p$-wave scattering length $a_p(E)$. Since the Rydberg electron's local kinetic energy—and thus the effective collision energy sampled by the scattering process—varies with internuclear distance, the value of $a_p(E)$ at the two nodal positions can differ significantly, particularly when the scattering length is near a shape resonance. As a result, the effective $p$-wave potential depth and the supported vibrational levels differ significantly between W$_1$ and W$_2$. Figure \ref{fig.4}(b) presents the measured spectrum of the ultralong-range Rydberg molecules. Two distinct molecular resonances are clearly resolved, corresponding to bound states supported by W$_1$ and W$_2$, respectively. The inset magnifies the shaded region, revealing additional vibrational excitations. The peaks labeled $\nu=0$ correspond to the ground vibrational state molecules ${}^3\Sigma(5s\text{-}35s)(\nu=0)$ of each well. The energy spacing and relative intensities of these vibrational series confirm that the two potential wells possess different curvatures and binding depths, consistent with the theoretical PEC shown in Fig. \ref{fig.4}(a). 

\subsection*{Stark effect of  \textit{p}-wave molecule}

To further characterize the features of the observed molecular states, we investigate their response to external static electric fields. Figure \ref{fig.5}(a) shows the molecular spectra measured under varying electric-field strengths. The spectral features located on the left and right sides of the spectrum correspond predominantly to the \textit{p}-wave and \textit{s}-wave dominated molecular states, respectively. As the electric field increases, both spectral lines exhibit quadratic frequency shifts, characteristic of the Stark effect. The magnitude and direction of the Stark shifts are almost the same for the two states, given the limited experimental precision of the present Stark-effect measurements. The extracted Stark shifts are plotted in Fig. \ref{fig.5}(b) as a function of the electric-field strength, and the data are well fitted by quadratic functions, confirming the dc Stark effect in the low-field regime. Representative spectra at selected electric-field strengths are displayed in Fig. \ref{fig.5}(c), clearly illustrating the evolution of both molecular features with the applied field. The observed quadratic Stark shifts, with polarizabilities quantitatively matching those of the bare \textit{n}\textit{S} Rydberg atom, are fully consistent with the weak-scattering limit of the Fermi pseudopotential.

\bibliography{ref}